\documentclass[review]{elsarticle}
\usepackage{lineno,hyperref}
\usepackage{epstopdf}
\usepackage{epsfig}
\modulolinenumbers[5]

\journal{Journal of \LaTeX\ Templates}

\def\dj{d\kern-0.4em\char"16\kern-0.1em}
\def\Dj{D\kern-1.65 ex\vrule height 0.95 ex depth -0.87 ex width 0.8ex\kern
0.78 ex}

\begin{document}

\begin{frontmatter}

\title{Dynamics of beryllium-7 specific activity in relation to meteorological variables, tropopause height,
teleconnection indices and sunspot number}

\author[FVM]{D.~Sarvan\corref{cor1}}
\ead{darko.sarvan@vet.bg.ac.rs}

\author[SF,IRACS]{{\Dj}.~Stratimirovi\'{c}}
\ead{dj.stratimirovic@gmail.com}

\author[DEIS,IRACS]{S.~Blesi\'{c}}
\ead{suzana.blesic@unive.it}

\author[IM]{V.~Djurdjevic}
\ead{vdj@ff.bg.ac.rs}

\author[FF]{V.~Miljkovi\'{c}}
\ead{vladimir.miljkovic@ff.bg.ac.rs}

\author[FVM,IRACS]{J.~Ajti\'{c}\corref{cor2}}
\ead{jelena.ajtic@vet.bg.ac.rs}

\cortext[cor1]{Corresponding author} \cortext[cor2]{Principal
corresponding author}

\address[FVM]{University of Belgrade, Faculty of Veterinary Medicine, Bulevar oslobo{\dj}enja 18, 11\,000 Belgrade, Serbia}
\address[SF]{University of Belgrade, Faculty of Dental Medicine, Dr Suboti\'{c}a 8, 11\,000 Belgrade, Serbia}
\address[IRACS]{Institute for Research and Advancement in Complex Systems, Zmaja od No\'{c}aja 8, 11\,000 Belgrade, Serbia}
\address[DEIS]{Department of Environmental Sciences, Informatics and Statistics, Ca'Foscari University of Venice, Mestre, Italy}
\address[IM]{University of Belgrade, Faculty of Physics, Institute of Meteorology, Studentski trg 12, 11\,000 Belgrade, Serbia }
\address[FF]{University of Belgrade, Faculty of Physics, Studentski trg 12, 11\,000 Belgrade, Serbia }

\begin{abstract}
The dynamics of the beryllium-7 specific activity in surface air
over 1987--2011 is analyzed using wavelet transform (WT) analysis
and time-dependent detrended moving average (tdDMA) method.
WT analysis gives four periodicities in the
beryllium-7 specific activity: one month, three months, one year,
and three years. These intervals are further used in tdDMA to
calculate local autocorrelation exponents for precipitation, tropopause height and teleconnection
indices. Our results show that these parameters share common
periods with the beryllium-7 surface concentration. tdDMA method
indicates that on the characteristic intervals of one year and
shorter, the beryllium-7 specific activity is strongly
autocorrelated. On the three-year interval, the beryllium-7
specific activity shows periods of anticorrelation, implying slow
changes in its dynamics that become evident only over a prolonged
period of time. A comparison of the Hurst exponents of all the
variables on the one- and three-year intervals suggest some
similarities in their dynamics. Overall, a good agreement in the
behavior of the teleconnection indices and specific activity of
beryllium-7 in surface air is  noted.
\end{abstract}


\end{frontmatter}


\section{Introduction}

Beryllium-7 (half-life 53.22 days) is a naturally occurring
radionuclide that is produced in the upper troposphere (around
30\,\%) and lower stratosphere (around 70\,\%) \cite{LalPet67}.
After formation $\mathrm{^7Be}$ attaches to fine aerosols, and its
residence time in the atmosphere is long \cite{Kochea96,
Tokea96,Dueea04, Heiea08}. The ensuing transport of the aerosols,
and therefore of the radionuclide, is governed by the atmospheric
circulation \cite{Gerea03, Criea06}.

Concentration of $\mathrm{^7Be}$ in any given location (altitude,
latitude, longitude) depends on several factors \cite{Feeea89}.
First, the source of the radionuclide is its production in the
higher layers of the atmosphere. Therefore, the production rate
influences the total amount of the radionuclide. Second, transport
can increase or decrease the radionuclide concentration
at a particular location, depending on abundance of
$\mathrm{^7Be}$ in the transported air masses. Finally, the rate
of the isotope removal influences its concentration in the
atmosphere.

The above mechanisms have been investigated. Monthly
$\mathrm{^7Be}$ specific activities in surface air are inversely
correlated with solar activity \cite{Gerea03,Canea04,Phamea11}.
Air masses originating in the upper troposphere and lower
stratosphere contain higher concentrations of $\mathrm{^7Be}$ than
surface air masses \cite{Bouea11}. Beryllium-7 can thus be used as
a stratospheric tracer, and has been investigated as an indicator
of exchange processes between the stratosphere and troposphere
\cite{Zanea03,Criea06}. Further, the $\mathrm{^7Be}$ concentration
maxima have been correlated with an enhanced vertical transport
and the intrusion of the stratospheric air masses across the
tropopause \cite{Todea99, Daiea05, Yoshimori05, Papas09}. A
positive correlation between the tropopause height and the
$\mathrm{^7Be}$ specific activity in surface air has been shown
\cite{Gerea01,Ioaea14}. Longitudinal and latitudinal distribution
of $\mathrm{^7Be}$ in the air has been noted
\cite{Feeea89,Kulea06,PerHol14,Herea15}, and it is in part
influenced by horizontal transport within the troposphere. Wet
deposition is the most significant mechanism of $\mathrm{^7Be}$
removal from the atmosphere \cite{Phamea11,PapIoa91,IoaPap06},
although different studies have shown no correlation or a negative
correlation of the $\mathrm{^7Be}$ specific activity with
precipitation \cite{Gerea03, Phamea11, Canea95, Aliea11, Chaoea12,
Garea12}.

To further explain the behavior of $\mathrm{^7Be}$ in surface air,
its relation to local climate variables, including (but not
limited to) temperature, atmospheric pressure, relative humidity,
and sunshine hours, has been extensively studied
\cite{PapIoa91,Todea99, Todea10, Ajtea08, Ajtea13, Phamea11,
Papea11, Carvalea13, Lozea13, Tosea14}. These studies, however,
did not include an analysis of $\mathrm{^7Be}$ relation with
large-scale atmospheric circulation.

Variability in atmospheric circulation is described by
teleconnection patterns, such as the North Atlantic Oscillation
(NAO), Arctic Oscillation (AO) and Pacific/North American
(PNA) \cite{WalGut81,BarLiv87}. These patterns are a measure of
pressure oscillations over different locations, and have been
shown to influence large-scale circulation \cite{Hurrell95,
AmbEA01, Wallace00}, which further reflects on local weather
conditions \cite{LeathersEA91, LeathersEA92, ThomWall98, HigEA02}.

An influence of NAO on $\mathrm{^7Be}$ has been implied
\cite{Criea06, Hedea06}, but only relatively recent studies have
focused on the $\mathrm{^7Be}$ specific activity in the air and
large-scale transport. For example, the abundance of
$\mathrm{^7Be}$ in Fennoscandia is not only influenced by NAO
\cite{Leppea10, Leppea12}, but the atmospheric conditions seem to
play a more important role than production \cite{LeppPaa13}.
Further, AO can modulate the stratosphere-troposphere exchange,
and as a consequence, the AO variability can explain a large part
of ozone variability in the lower troposphere over North America
\cite{LamHes04}. This finding is also relevant for $\mathrm{^7Be}$
which is, along with ozone, transported from the stratosphere into
troposphere.

To summarize, there have been a number of studies on the
$\mathrm{^7Be}$ specific activity in surface air and its relation
with local meteorological conditions, sunspot number, and
tropopause height, and somewhat fewer studies on the influence of
large-scale atmospheric transport. Most of these studies looked
into linear relationship between the $\mathrm{^7Be}$ specific
activity and a set of chosen variables. However, there have been
no in-depth statistical analysis encompassing meteorological
variables, tropopause height, sunspot number and teleconnection
indices (which quantify large-scale transport). The goal of our
investigation is to look into common periodicities of the
mentioned variables, whose existence could help to understand a
relationship between the variables, even if it may not be linear
in its nature. Two statistical analysis methods are used to
investigate the dynamics of the $\mathrm{^7Be}$ surface
concentration: wavelet transform analysis and time-dependant
Hurst exponent method.

\section{Data}

Time series of 11 measured variables were
analyzed. The $\mathrm{^7Be}$ specific activity in surface air,
five meteorological variables, and the tropopause height were of
local character -- the data were recorded in Helsinki, Finland
($\mathrm{60.21\,^{\circ}N}$; $\mathrm{25.06\,^{\circ}E}$; 12 m
a.s.l). On the other hand, three teleconnection indices and sunspot number quantify hemispheric circulation, and sun
activity, respectively. This set of variables was chosen in
attempt to include as many as possible factors potentially
influencing the $\mathrm{^7Be}$ specific activity in surface air,
but this choice was limited by the number of meteorological
variables available for Helsinki, and by the temporal resolution
of the available teleconnection indices. The length of the investigated time series differed, and the
start and end date of the analysis were chosen to coincide with
the available $\mathrm{^7Be}$ specific activity data -- from 1
January 1987 to 31 December 2011, thus spanning 25 years.

\paragraph{Beryllium-7 specific activity in surface air} The analyzed data are a subset of the Radioactivity
Environmental Monitoring Database (REMdb) supported by REM group
from the Institute of Transuranium Elements, of the DG Joint
Research Centre (JRC). The $\mathrm{^7Be}$ data prior to 2007 stored in the
REMdb is public, and an access to the data
over the 2007--2011 period can be granted only after
explicit request. More information on the REMdb can be found on
its web page (https://rem.jrc.ec.europa.eu/) and in
\cite{Herea15,Herea16,Hernandezetal16}. The measurements conducted in Helsinki represent the largest set
with more than 4 000 data points over 1987--2011. The sampling
frequency of the measurements varied: prior to 1999, the
measurements were taken mostly once a week, while the subsequent
measurements were performed daily or once in two days.

\paragraph{Meteorological variables}The meteorological data, consisting of mean, minimum and maximum
temperature, atmospheric pressure, and precipitation, were
obtained from the European Climate Assessment \& Dataset (ECA\&D)
\cite{Klein_Tankea02}. The series consisted of daily data.

\paragraph{Tropopause height}Tropopause height was calculated following the procedure given in \cite{Ioaea14}.
Input data for the calculations were taken form the NCEP/NCAR
reanalysis \cite{NCEP/NCAR96}.  In the procedure, an extrapolation
of isobaric heights above and below the tropopause to the
tropopause pressure was performed using the hydrostatic
approximation. The averaged value of the two extrapolated values
was then taken as the height of the tropopause. The calculations of
the tropopause height were performed for each day of the
investigated period.

\paragraph{Teleconnection indices}The daily values of three teleconnection indices of large-scale atmospheric circulation:
North Atlantic Oscillation, Arctic Oscillation, and
Pacific/North American were obtained from the data archive
of the United States National Oceanic and Atmospheric
Administration's Climate Prediction Center
(http://www.cpc.ncep.noaa.gov visited on 16 April 2015).

\paragraph{Sunspot number}The daily sunspot data were obtained from the SIDC-team (World Data Center for the Sunspot Index,
Royal Observatory of Belgium, Monthly Report on the International
Sunspot Number, online catalogue of the sunspot index:
http://www.sidc.be/sunspot-data/1987-2011).

\section{Calculations}
Two statistical methods were applied in the analysis of the
$\mathrm{^7Be}$ specific activity dynamics: wavelet transform
analysis and time-dependant Hurst exponent method.

\subsection{Wavelet analysis}

Wavelet transform (WT) spectral analysis of the $\mathrm{^7Be}$
time series was used to look into existence of periodic or
quasi-periodic cycles in the data \cite{BraSte98,Kikuea09} and to
assess the overall statistical behavior of the $\mathrm{^7Be}$
series \cite{TorCom98}. The existence of cycles was then compared
across the chosen datasets to find their common periodicities.

Wavelet transform analysis is commonly used as a tool to
investigate time series that contain nonstationarities on a number
of different frequencies \cite{Lewea07}. The WT procedure that we used is described in \cite{TorCom98,StratimirovicEA01}.
In this paper, a set of Derivatives
of Gaussian (DOG) wavelets of the tenth order was used to
calculate WT coefficients. The calculated
wavelet spectra represent variations of the analyzed signals on different
time scales, and show increased values for the events occurring at
a characteristic time scale. To detect those characteristic
scales, a standard peak analysis was performed by searching the
maximum and saddle (for hidden peaks) points in the WT
power spectra of the investigated variables.

\subsection{Time-dependant Hurst exponent}

Centered detrended moving average (cDMA) technique operates
through an estimate of a generalized variance of the long-range
correlated series $y(i)$ around the moving average
$\tilde{y}_n(i)$:
\begin{equation}\label{DMA1}
    \tilde{y}_n(i)=\frac{1}{n}\sum_{k=-\frac{n}{2}}^{k=\frac{n}{2}-1}y(i-k)\,,
\end{equation}
where $n$ is the width  of the moving average window. The time
series $y(i)$ is detrended by subtracting the local trend
$\tilde{y}_n(i)$, and for a given window width $n$, the
characteristic size of fluctuation for detrended time series is
calculated by:
\begin{equation}\label{DMA2}
    \sigma_{cDMA}(n)=\sqrt{\frac{1}{N_{max}-n}
        \sum_{i=\frac{n}{2}}^{N_{max}-\frac{n}{2}}[y(i)-\tilde{y}_n(i)]^2}\,.
\end{equation}
Function $\sigma_{cDMA}(n)$ is calculated for different moving
window widths, $n\in[\frac{n}{2}, N_{max}-\frac{n}{2}]$, where
$N_{max}$ is the length of the entire series. An increase in the
window width $n$ increases the function $\sigma_{cDMA}(n)$.

When the analyzed time series follows a scaling law, the cDMA
function is of a power-law type, i.e. $\sigma_{cDMA}(n) \propto
n^H$, where $H$ is Hurst exponent which is related to the
correlation properties of $y(i)$. When $ 0.5 <H <1 $, the series
$y(i)$ has a positive long-range correlation, or persistence; when
$ 0 <H <0.5 $, the series has a long-range negative correlation,
or anti-persistence; and when $ H = 0.5$, the series can be
described as an uncorrelated Brownian process.

Local complexity in our data sets was investigated using the
time-dependent DMA (tdDMA) algorithm \cite{Carea04} in the following manner. First, cDMA algorithm was applied on the subset of data at the intersection of
the time series signal and a sliding window with a width $N_s$,
which moved along the series with a step $\delta_s$. The scaling
exponent $H$ was then calculated for each subset, following the cDMA
procedure described above, and a sequence of local, time-dependent
Hurst exponents was obtained. The minimum size of each subset
$N_{min}$ was defined under a condition that the scaling law
$\sigma_{cDMA}(n) \propto n^H$ holds in the subset, while the
accuracy of the technique was achieved with an appropriate choice
of $N_{min}$ and $\delta_{min}$ \cite{Conea13}.

In our tdDMA analysis, window widths of up to $N_s$ = 3265 were chosen, with the step
$\delta_s=1$. The scaling features of the chosen variables
were studied on four characteristic periods: one month, three months, one year, and three years.
These periods enclosed the $\mathrm{^7Be}$ specific activity spectral peaks obtained by the WT analysis.

\section{Results and Discussion}

The wavelet spectra for the $\mathrm{^7Be}$ data series were
calculated in the range of 15--1500 points, corresponding to the
time span of ten days to four years. This range was chosen having
in mind the length of the time series analyzed (daily records in
25 years), so that we can obtain statistically relevant results
\cite{PengEA93}. 

Augmented Dickey-Fuller unit-root test was performed on the $\mathrm{^7Be}$ data series. The test showed stationarity of the data series. 

\subsection{Characteristic periods}

\begin{figure}[h!]
\includegraphics[scale=0.35]{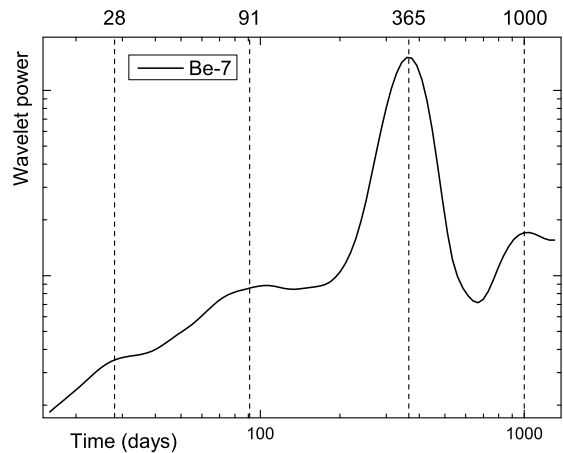}%
\caption {Wavelet spectrum of the $\mathrm{^7Be}$ specific
activity in surface air, Helsinki. \label{Fig1}
}%
\end{figure}

Four characteristic periods (corresponding to the time coordinates
of the local maxima) were detected in the $\mathrm{^7Be}$ spectrum
(Fig.\,\ref{Fig1}): the peak around 30 days indicating a monthly
cycle; the peak around 90 days indicating a seasonal cycle; the
peak around 360 days indicating an annual cycle; and the peak
around 1\,000 days implying a longer cycle of around three
years (triennial cycle). A seasonal periodicity has already been
observed in the behavior of $\mathrm{^7Be}$ \cite{YamamotoEA06,
Leppea12}. A periodicity of 45--90 days in the $\mathrm{^7Be}$
wavelet spectrum has also been reported before \cite{LeppPaa13},
as an intermittent period connected to changes in teleconnection
indices. Similarly, a periodicity of $\sim2.5$ years in the
$\mathrm{^7Be}$ surface concentrations has already been noted
\cite{Gerea03}.

\begin{figure}[h!]
\includegraphics[scale=0.35]{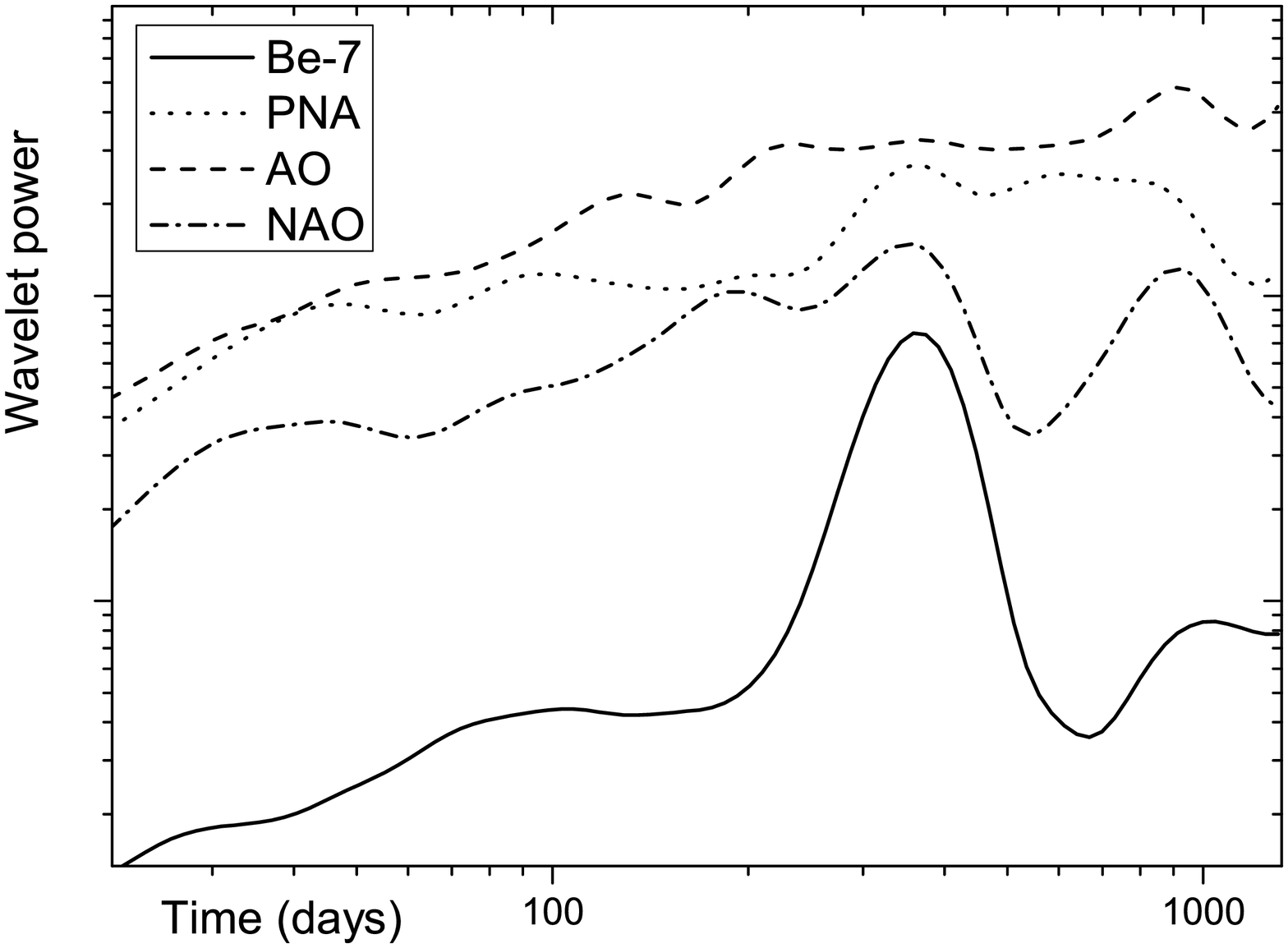}%
\caption {Wavelet transform spectra of the $\mathrm{^7Be}$
specific activity and teleconnection indices: Pacific/North
American (PNA), Arctic Oscillation (AO), and
North Atlantic Oscillation (NAO). \label{Fig2} }
\end{figure}

Figure 2 shows a comparison of the $\mathrm{^7Be}$ and
teleconnection indices WT spectra. In general, the teleconnection
indices displyed a distinct annual period, and their visible peaks
were positioned close to the seasonal and triennial
 $\mathrm{^7Be}$ periods (Fig.\,\ref{Fig2}). The spectral changes of the NAO and PNA indices were particularly
consistent with the changes in the $\mathrm{^7Be}$ spectrum (with
correlation coefficients of 0.7 and 0.6, respectively).

\begin{figure}[h!]
\includegraphics[scale=0.35]{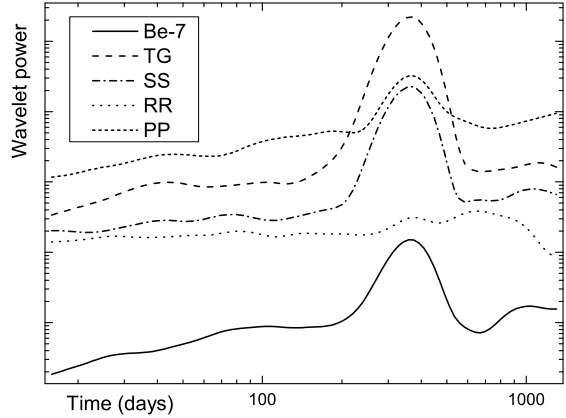}%
\caption {Wavelet transform spectra of the $\mathrm{^7Be}$
specific activity and meteorological data: mean daily temperature
(TG), sunspot number (SS), precipitation (RR), and atmospheric
pressure (PP). \label{Fig3} }
\end{figure}

The meteorological data also showed a distinct annual peak, with additional
peaks not coinciding with the $\mathrm{^7Be}$
periods (Fig.\,\ref{Fig3}). The WT spectra for all temperature records,
atmospheric pressure and sunspot number correlated well and
significantly with the $\mathrm{^7Be}$ spectrum (with correlation
coefficients larger than 0.9). The only time series not following
the strong correlation pattern was precipitation. These findings
could indicate that in general, the atmospheric conditions
strongly influence the $\mathrm{^7Be}$ annual cycle, while
temperature, atmospheric pressure, and sunspot number possibly
have a bigger influence on the overall $\mathrm{^7Be}$ variations.
The triennial $\mathrm{^7Be}$ cycle seems to be influenced by a combination of
atmospheric conditions and teleconnections, which is in partial
agreement with previous studies \cite{Gerea03, Leppea12, AngKor64}.

\begin{figure}[h!]
\includegraphics[scale=0.35]{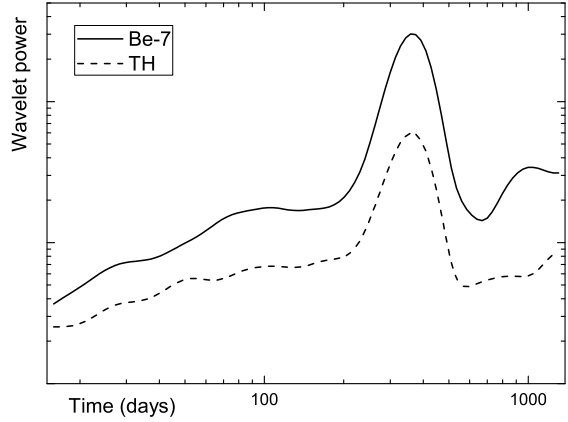}%
\caption {Wavelet transform spectra of the $\mathrm{^7Be}$
specific activity and tropopause hight (TH). \label{FigTH} }
\end{figure}

Finally, the data for tropopause height showed a good overall
agreement with the $\mathrm{^7Be}$ cycles (with correlation
coefficient larger than 0.9). As depicted in Fig.\,\ref{FigTH},
the tropopause height records displayed a distinct annual peak,
together with less evident monthly and seasonal peaks. The
tropopause height gave a multiannual peak indicating a period
somewhat shorter than the triennial $\mathrm{^7Be}$ peak. This
tropopause height periodicity of $\sim$2.5 years has been
identified previously \cite{AngKor64}. A positive correlation
between the tropopause height and the $\mathrm{^7Be}$ specific
activity in surface air was observed by \cite{Gerea01,Ioaea14,Hernandezetal16}.

\subsection{Temporal dynamics of the $\mathrm{^7Be}$ surface concentration}

The $\mathrm{^7Be}$ periods found in WT analysis were further used
in tdDMA to investigate features of long-range dependence in the
datasets. Specifically, the values of the Hurst exponents
calculated on these intervals describe autocorrelation behavior of
each investigated variable. In addition, a comparison of the local
Hurst exponents of different variables on a given characteristic
scale could unveil similarities in their dynamics.

\begin{figure}[h!]
\includegraphics[scale=0.35]{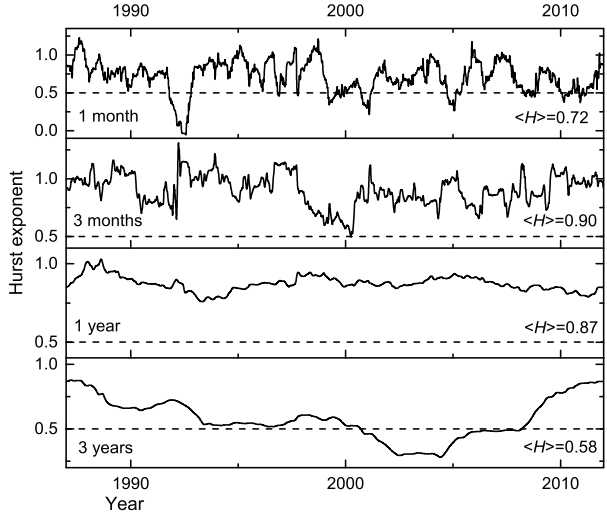}%
\caption {Local Hurst exponents and their mean values for the $\mathrm{^7Be}$
specific activity on its characteristic periods. \label{FigBe7} }
\end{figure}

Figure\,\ref{FigBe7} shows the local Hurst exponents for the
$\mathrm{^7Be}$ specific activity on its characteristic periods.
On the shortest characteristic interval of one month, the
exponents implied a strong autocorrelation, with the mean Hurst
exponent of 0.72. With an increase in the characteristic period to three months, the mean
Hurst exponent increased to 0.90, and this same pattern was repeated on the interval
of one year -- the mean Hurst exponent was 0.87, with no
anticorrelation bouts. A significant change in the behavior of the
local Hurst exponents occurred on the interval of three years, where we found a multiyear period with the exponents close to 0.5,
followed by an anticorrelated regime. The mean Hurst exponent for
this three-year period decreased to 0.58 (Fig.\,\ref{FigBe7}).

The local Hurst exponents on different characteristic intervals
suggest that the temporal changes in the $\mathrm{^7Be}$ specific
activity are most likely slow. Namely, the high values of the exponents on shorter
time-scales of one and three months, and even one year, imply
almost identical behavior in the $\mathrm{^7Be}$ records. For
example, with the mean Hurst exponent of 0.90, within every
trimester, the type of changes (i.e., an increase or decrease in the measured values) were almost the same as the type of changes within the preceding and subsequent trimester. Similarly, the differences in the type of changes from one year to another also seemed negligible. Only over a prolonged period of time, which was three
years in the case of the $\mathrm{^7Be}$ specific activity, we observed a variation in the $\mathrm{^7Be}$ dynamics: the highly correlated regime apparently shifted to slightly correlated, even anticorrelated behavior. This would suggest an existence of a crossover in the radionuclide's behavior on large (multiyear) scales that is maybe masked by the presence of a one-year peak in the $\mathrm{^7Be}$ wavelet spectrum. 

\subsection{Comparative analysis of the variables' dynamics}

\begin{figure}[h!]
\includegraphics[scale=0.5]{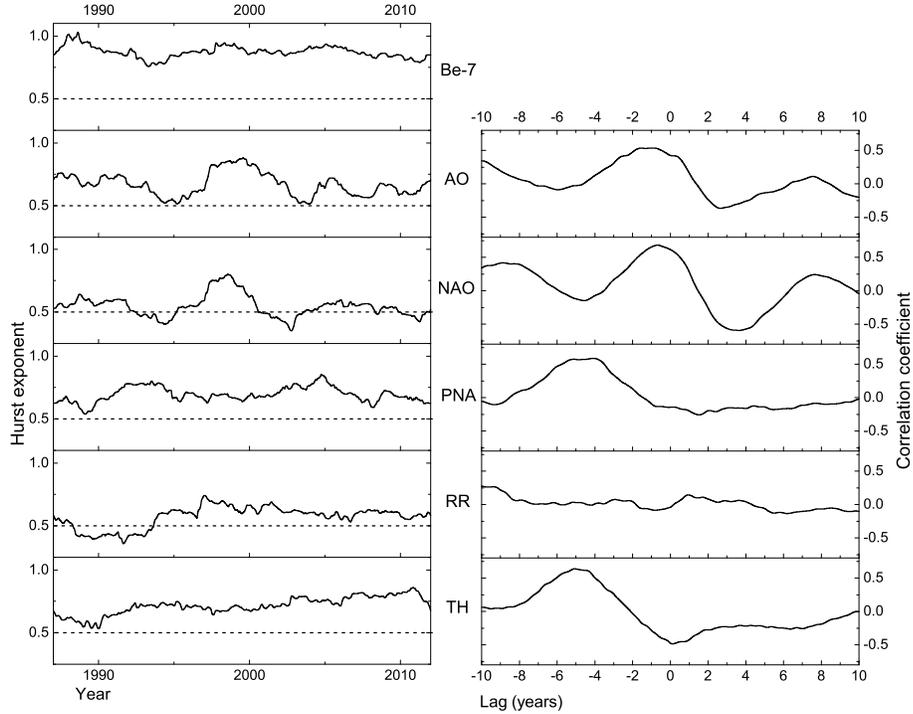}%
\caption {Left -- local Hurst exponents on the one-year interval
for (top to bottom): the $\mathrm{^7Be}$ specific activity, Arctic
Oscillation, North Atlantic Oscillation, Pacific/North American,
precipitation, and tropopause height, over 1987--2011. Right --
correlation function between the local Hurst exponents of the
$\mathrm{^7Be}$ specific activity and the corresponding variables.
\label{Fig4} }
\end{figure}

To further analyze the observed behavior and possibly find a source of crossover in the $\mathrm{^7Be}$ dynamics, we compared
its local Hurst exponents on the one- and three-year intervals
with the exponents of the other investigated variables
(Figs.\,\ref{Fig4} and \ref{Fig5}). Similarities in the temporal
evolution of these Hurst exponents are described by the
correlation functions also shown in Figs.\,\ref{Fig4} and
\ref{Fig5}, with some characteristic values given in
Table\,\ref{Table1}.

\subsubsection{Dynamics over one year}

On the one-year interval, the variables were autocorrelated, apart
from NAO and precipitation over short periods (Fig.\,\ref{Fig4}).
The minima of the correlation functions occurred when the
$\mathrm{^7Be}$ specific activity was correlated with the
corresponding variable's values recorded later in time
(Table\,\ref{Table1}). On the other hand, the maximum correlations
showed a moderate to strong correlation (except precipitation)
which occurred when the $\mathrm{^7Be}$ specific activity was
correlated with the corresponding variable's values recorded
earlier in time. The strength of anticorrelation was less than the
strength of correlation (i.e., the absolute value of the minimum
was less than the absolute value of the maximum). These
correlation coefficients could imply that changes in the
teleconnection indices reflected on the changes in the
$\mathrm{^7Be}$ specific activity with a varying time lag: from
less than a year for AO and NAO, to $\sim$4 years for PNA.

\begin{table}[]
\centering \caption{The correlation value for time lag equal 0,
maximum and minimum of the correlation functions given in
Figs.\,\ref{Fig4} and \ref{Fig5}. The time lags for which the
minimum and maximum are reached are given in years. When time lag
is positive, the $\mathrm{^7Be}$ specific activity is correlated
with the corresponding variable's values recorded earlier in time
("$\mathrm{^7Be}$ is in the future"), and vice versa.}
\label{Table1}
\resizebox{\textwidth}{!}{
\begin{tabular}{|l|c|c|c|c|c|c|} \hline
Period & \multicolumn{3}{c|}{One year} & \multicolumn{3}{c|}{Three
years}\\\hline
 & corr. coeff. & maximum & minimum
& corr. coeff. &
maximum & minimum           \\
\raisebox{1.5ex}[0pt]{Variable} & lag=0 & (lag in years) & (lag in
years) & lag=0 & (lag in years) & (lag in
 years)\\\hline
      & & 0.54  & -0.37  &  & 0.19  & -0.72  \\
\raisebox{1.5ex}[0pt]{AO} & \raisebox{1.5ex}[0pt]{0.42} & (-0.9) &
(2.7) & \raisebox{1.5ex}[0pt]{-0.02} & (4.3) & (-3.6)\\\hline
      & & 0.68  & -0.60  &  & 0.30  & -0.71 \\
\raisebox{1.5ex}[0pt]{NAO} & \raisebox{1.5ex}[0pt]{0.61} &(-0.7) &
(3.6)  & \raisebox{1.5ex}[0pt]{-0.29} & (10) & (9.3) \\\hline
      & & 0.59  & -0.26  &  & 0.56  & -0.51 \\
\raisebox{1.5ex}[0pt]{PNA} & \raisebox{1.5ex}[0pt]{-0.14} &(-4.1)
& (1.5) & \raisebox{1.5ex}[0pt]{0.56} & (0) & (-4.8) \\\hline
      & & 0.27  & -0.14  &  & 0.45  & -0.86 \\
\raisebox{1.5ex}[0pt]{RR} & \raisebox{1.5ex}[0pt]{-0.04} &(-10) &
(6.2) & \raisebox{1.5ex}[0pt]{-0.52} & (10) & (3.4)\\\hline
      & & 0.64  & -0.49  &  & 0.42  & -0.65 \\
\raisebox{1.5ex}[0pt]{TH} & \raisebox{1.5ex}[0pt]{-0.47} &(-5.1) &
(0.1) & \raisebox{1.5ex}[0pt]{0.09} & (-10) & (6.1) \\\hline
\end{tabular}
}
\end{table}

It is worth noting here that the extremes given in
Table\,\ref{Table1}, which are found with the time lag of 10
years, should be taken with caution, as that time lag is at the
very limit of the analyzed time window. This was the case with
precipitation on the one-year interval, and NAO, precipitation and
tropopause height on the three-year interval.

\subsubsection{Dynamics over three years}

\begin{figure}[h!]
\includegraphics[scale=0.5]{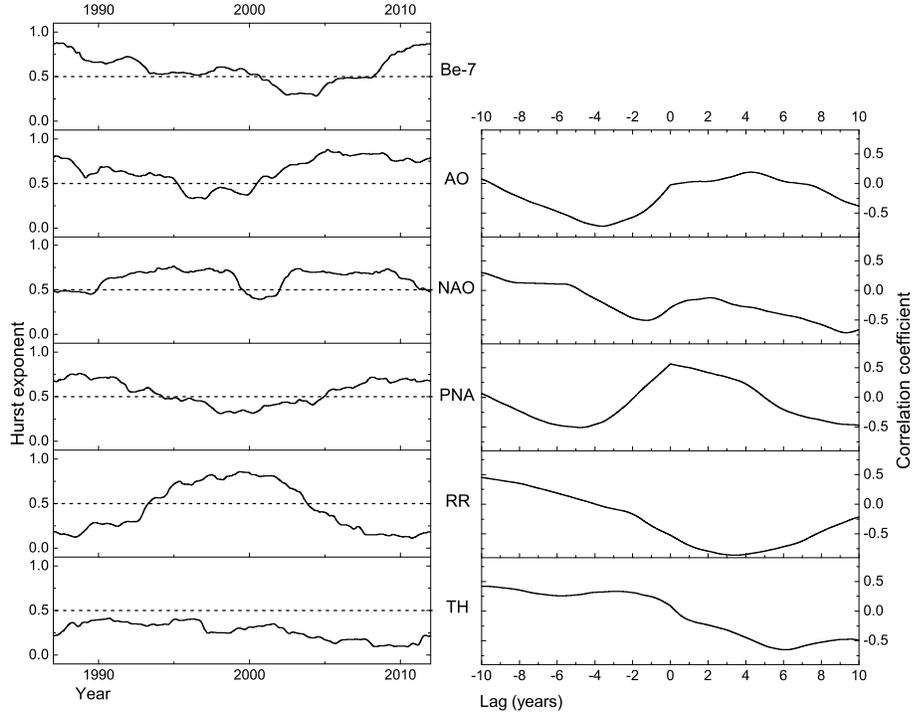}%
\caption {Same as Fig.\,\ref{Fig4} on the three-year
interval.\label{Fig5} }
\end{figure}

The long-range autocorrelations (on the three-year interval) give
insight into the complexity of the variables' behavior
(Fig.\,\ref{Fig5}). While on the one-year interval, there were no
significant variations in the type of changes in the $\mathrm{^7Be}$ specific activity from one year to the next, the temporal evolution was different on the
three-year interval. Although the radionuclide's concentration showed strong
autocorrelation prior to 1990 and after 2010, there were
significant differences during the period in-between. The transition in
the correlation mode, from correlated to anticorrelated, occurred
in all of the variables, except the tropopause height which was
anticorrelated throughout the 1987--2011 period
(Fig.\,\ref{Fig5}).

During 1993--2005, the local $\mathrm{^7Be}$ Hurst exponent
decreased, which may suggest that over this period there was a
weakened effect of some global parameters, which have an impact on
the dynamics of the $\mathrm{^7Be}$ specific activity. Over this
period, AO, PNA and precipitation changed the mode, i.e. from
correlated to anticorrelated, and vice versa. The transition of
the AO and PNA teleconnection indices into the anticorrelated
regime, which happened in $\sim$1995, could have driven the
$\mathrm{^7Be}$ specific activity shift into an anticorrelated
regime. The correlation in the $\mathrm{^7Be}$ series was first
reduced between 1995 and 2000, and the actual transition in the
correlation mode (from correlated to anticorrelated) occurred in
2000. Further evolution of the Hurst exponents showed that AO
reversed back into the correlated regime in 2000, but the PNA
reversal took another five years. The transition back into the
correlated mode of the $\mathrm{^7Be}$ specific activity occurred
later still, in 2008.

The maximum and minimum values of the correlation functions
(Fig.\,\ref{Fig5} and Table\,\ref{Table1}) show that in contrast
to the one-year interval, anticorrelation between the Hurst exponents of the investigated variables and the $\mathrm{^7Be}$ specific activity was stronger
than their correlation. Further, the minima for the AO and PNA
correlation functions occurred for a time lag of 4--5 years, which
agrees with the aforementioned delay with which the mode reversals
of the $\mathrm{^7Be}$ Hurst exponents followed the shifts in the
teleconnections' autocorrelations (Fig.\,\ref{Fig5}).

The slow changes in the dynamics of the $\mathrm{^7Be}$ specific
activity could be interconnected with the changes in the
teleconnections. In one scenario of the interconnection, a direct
influence of the indices on the $\mathrm{^7Be}$ specific activity
could be assumed. This scenario would then suggest a relatively
long response time (4--5 years) in which the correlation changes
in large-scale atmospheric circulation were mirrored by the
correlation changes in the radionuclide activity. On the other
hand, if the teleconnection indices and $\mathrm{^7Be}$ specific
activity had a common driving mechanism, then the changes would
first occur in the behavior of AO, then PNA and $\mathrm{^7Be}$ specific
activity (Fig.\,\ref{Fig5}). The correlation mode shifts for AO
and PNA were not only earlier, but also faster (note a prolonged
period in the second half of the 1990s during which the
$\mathrm{^7Be}$ Hurst exponent hovered close to 0.5). This slow
reaction of the $\mathrm{^7Be}$ specific activity to a driving
mechanism would result from additional forces that dampen the
influence and thus prolong the response time. Another potential interconnection between the variables is a combination of the two scenarios, encompassing a common
driving mechanism and an influence that teleconnections exert on
the $\mathrm{^7Be}$ specific activity in surface air.

A possible explanation for the observed change in the correlation
of the variables (Fig.\,\ref{Fig5}) could lie in the solar
activity during 1993--2005. This period encompassed the end of
solar cycle 22 (1986--1996) and most of solar cycle 23
(1996--2008). In contrast to solar cycle 22 with high solar
activity, solar cycle 23 was noticeably smaller \cite{NASA},
and the changes in correlation modes could reflect those changes
in solar activity. It is worth noting here that over 1993--2005,
the decrease in the correlation of the $\mathrm{^7Be}$ series
was not associated with a decrease in the measured $\mathrm{^7Be}$ specific activity. A major influence of solar forcing on the $\mathrm{^7Be}$ surface concentration for timescales longer than one year has been noticed by \cite{Talposetal05}.

It is also interesting to note that the precipitation Hurst exponents on the three-year
period (Fig.\,\ref{Fig5}) showed a decrease in anticorrelation
from 1987 to 1994, and an increase in anticorrelation from 2006 to
2011, just as the Hurst exponents of the $\mathrm{^7Be}$ specific activity exhibited the
opposite behavior. Over the period in-between, 1994--2006, precipitation was
autocorrelated while the $\mathrm{^7Be}$ Hurst exponent hovered around
0.5 and then shifted into anticorrelation (Fig.\,\ref{Fig5}).
Further, very strong anticorrelation between the precipitation and
$\mathrm{^7Be}$ Hurst exponents of -0.86 was noted with a time lag
of approximately three years (Table\,\ref{Table1}).

Finally, some caveats in our analysis should be mentioned. The
availability of the $\mathrm{^7Be}$ data limited the analysis to
time periods from ten days to four years, not allowing
investigation of periodicities on shorter time scales, such as the
cycles found in a wavelet analysis of daily $\mathrm{^7Be}$
concentrations \cite{Kikuea09}. It also limited our insight into
longer, multiannual or interdecadal, periodicities reported in
\cite{Leppea10}. Moreover, WT spectral analysis is a linear method
which detects linear combinations of characteristic frequencies
unless their magnitude is very small \cite{BraSte98}. Further
studies should aim at improvements that look into recognition of
these linear combinations, and their separation from true
characteristic frequencies inherent to the data. The future
modelling efforts could possibly profit from the results of the
comparison of the WT spectra of the $\mathrm{^7Be}$ specific
activity, meteorological parameters, and teleconnection indices,
and use our findings for re-calibration of modelling strategies.

Further, in our analysis an emphasis was given to an
interrelationship between the $\mathrm{^7Be}$ specific activity
and other parameters. A mutual linkage between the parameters was
not discussed, although our results offered some ready
conclusions. For example, on the short characteristic intervals of
one and three months, all teleconnection indices were
autocorrelated (not shown), while periods of anticorrelation first
occurred in NAO on the one-year interval (Fig.\,\ref{Fig4}) and
then in AO and PNA on the three-year interval (Fig.\,\ref{Fig5}).
These results imply a very similar behavior between the AO and PNA
teleconnections and the $\mathrm{^7Be}$ specific activity,
including slow changes which become evident only on the three-year
interval.

\section{Conclusions}

Wavelet transform spectral analysis revealed the existence of four
characteristic periods in the $\mathrm{^7Be}$ specific activity: a
one-month cycle, a three-month cycle, an annual cycle, and a
triennial cycle. Our results implied that the tropopause height,
atmospheric conditions (especially temperature), atmospheric
pressure, and the sunspot number, significantly influence the
overall $\mathrm{^7Be}$ spectral behavior. The seasonal period
also seems to be influenced by large-scale atmospheric
circulation, particularly by the NAO and PNA indices, and by the
tropopause height, which seems to also influence the monthly
$\mathrm{^7Be}$ period. The triennial $\mathrm{^7Be}$ cycle is
likely to be influenced by a combination of atmospheric conditions
and teleconnections. In our data, the characteristic time periods
of the NAO and PNA patterns seemed to match the periods of the
radionuclide's activity better than AO.

Local Hurst exponent analysis offered further insight into the
complex dynamic of the $\mathrm{^7Be}$ specific activity in
surface air and its relationship to dynamics of meteorological
parameters and teleconnection indices. Comparisons were made on
the $\mathrm{^7Be}$ specific activity characteristic periods of
one year and three years. On the time period of one year and
shorter, the $\mathrm{^7Be}$ specific activity was strongly
autocorrelated. The longest, three-year interval, showed some
periods of anticorrelation implying that changes in the dynamics
of this radionuclide are slow and evident only on the scale of
three years. Similarities in the overall pattern of the Hurst
exponents on the four characteristic intervals suggest a good
agreement in the AO, PNA and $\mathrm{^7Be}$ specific activity
behavior.

\section*{Acknowledgements}
This paper was realized within the following projects: "Advanced
analytical, numerical and methods of analysis of fluid mechanics
and complex systems" (No. OI174014), "Phase transitions and
characterization of inorganic and organic systems" (No. OI171015),
and "Climate changes and their influence on the environment:
impacts, adaptation and mitigation" (No. III43007), financed by
the Ministry of Education, Science and Technological Development
of the Republic of Serbia (2011-2016). Suzana Blesi\'c has received funding from the European Union’s Horizon 2020 research and innovation programme under the Marie Sklodowska-Curie grant agreement No. 701785. The authors would like to
thank the REM group for provision of the beryllium-7 specific
activity measurements from the REM data base (REMdb at the
Institute of Transuranium Elements, REM group, DJ JRC Ispra site,
European Commission).


\begin{thebibliography}{}
\expandafter\ifx\csname url\endcsname\relax
  \def\url#1{\texttt{#1}}\fi
\expandafter\ifx\csname urlprefix\endcsname\relax\def\urlprefix{URL }\fi
\expandafter\ifx\csname href\endcsname\relax
  \def\href#1#2{#2} \def\path#1{#1}\fi

\end{thebibliography}


\section*{References}

\begin{thebibliography}{10}
\expandafter\ifx\csname url\endcsname\relax
  \def\url#1{\texttt{#1}}\fi
\expandafter\ifx\csname urlprefix\endcsname\relax\def\urlprefix{URL }\fi
\expandafter\ifx\csname href\endcsname\relax
  \def\href#1#2{#2} \def\path#1{#1}\fi

\bibitem{LalPet67}
D.~Lal, B.~Peters, Cosmic ray produced radioactivity on the {E}arth, in:
  K.~Sitte (Ed.), Kosmische Strahlung II / Cosmic Rays II, Vol. 9 / 46 / 2 of
  Handbuch der Physik / Encyclopedia of Physics, Springer Berlin Heidelberg,
  1967, pp. 551--612.
\newblock \href {http://dx.doi.org/10.1007/978-3-642-46079-1_7}
  {\path{doi:10.1007/978-3-642-46079-1_7}}.

\bibitem{Kochea96}
D.~M. Koch, D.~J. Jacob, W.~C. Graustein, Vertical transport of tropospheric
  aerosols as indicated by $\mathrm{^7Be}$ and $\mathrm{^{210}Pb}$ in a
  chemical tracer model, J. Geophys. Res. 101~(D13) (1996) 18651--18666.
\newblock \href {http://dx.doi.org/10.1029/96JD01176}
  {\path{doi:10.1029/96JD01176}}.

\bibitem{Tokea96}
T.~Tokieda, K.~Yamanaka, K.~Harada, S.~Tsunogai, Seasonal variations of
  residence time and upper atmospheric contribution of aerosols studied with
  {P}b-210, {B}i-210, {P}o-210 and {B}e-7, Tellus B 48~(5).
\newblock \href {http://dx.doi.org/10.1034/j.1600-0889.1996.t01-4-00006.x}
  {\path{doi:10.1034/j.1600-0889.1996.t01-4-00006.x}}.

\bibitem{Dueea04}
C.~{Due\~{n}as}, M.~Fern\'{a}ndez, J.~Carretero, E.~Liger, S.~{Ca\~{n}ete},
  Long-term variation of the concentrations of long-lived {R}n descendants and
  cosmogenic $\mathrm{^7Be}$ and determination of the {MRT} of aerosols, Atmos.
  Environ. 38~(9) (2004) 1291--1301.
 
\bibitem{Heiea08}
U.~Heikkil\"{a}, J.~Beer, V.~Alfimov, Beryllium-10 and beryllium-7 in
  precipitation in {D}\"{u}bendorf (440 m) and at {J}ungfraujoch (3580 m),
  {S}witzerland (1998-2005), J. Geophys. Res. 113~(D11).
\newblock \href {http://dx.doi.org/10.1029/2007JD009160}
  {\path{doi:10.1029/2007JD009160}}.

\bibitem{Gerea03}
E.~Gerasopoulos, C.~Zerefos, C.~Papastefanou, P.~Zanis, K.~O'{B}rien,
  Low-frequency variability of beryllium-7 surface concentrations over the
  {E}astern {M}editerranean, Atmos. Environ. 37~(13) (2003) 1745--1756.

\bibitem{Criea06}
P.~Cristofanelli, P.~Bonasoni, L.~Tositti, U.~Bonaf\`{e}, F.~Calzolari,
  F.~Evangelisti, S.~Sandrini, A.~Stohl, A 6-year analysis of stratospheric
  intrusions and their influence on ozone at {M}t. {C}imone (2165 m above sea
  level), J. Geophys. Res. 111~(D3) (2006), d03306.
\newblock \href {http://dx.doi.org/10.1029/2005JD006553}
  {\path{doi:10.1029/2005JD006553}}.

\bibitem{Feeea89}
H.~W. Feely, R.~J. Larsen, C.~G. Sanderson, Factors that cause seasonal
  variations in {B}eryllium-7 concentrations in surface air, J. Environ.
  Radioact. 9~(3) (1989) 223--249.

\bibitem{Canea04}
F.~Cannizzaro, G.~Greco, M.~Raneli, M.~Spitale, E.~Tomarchio, Concentration
  measurements of $\mathrm{^7Be}$ at ground level air at {P}alermo,
  Italy--comparison with solar activity over a period of 21 years, J. Environ.
  Radioact. 72~(3) (2004) 259--271.

\bibitem{Phamea11}
M.~K. Pham, M.~Betti, H.~Nies, P.~P. Povinec, Temporal changes of
  $\mathrm{^7Be}$, $\mathrm{^{137}Cs}$ and $\mathrm{^{210}Pb}$ activity
  concentrations in surface air at {M}onaco and their correlation with
  meteorological parameters, J. Environ. Radioact. 102~(11) (2011) 1045--1054.

\bibitem{Bouea11}
L.~Bourcier, O.~Masson, P.~Laj, J.~Pichon, P.~Paulat, E.~Freney, K.~Sellegri,
  Comparative trends and seasonal variation of $\mathrm{^7Be}$,
  $\mathrm{^{210}Pb}$ and $\mathrm{^{137}Cs}$ at two altitude sites in the
  central part of {F}rance, J. Environ. Radioact. 102~(3) (2011) 294--301.

\bibitem{Zanea03}
P.~Zanis, E.~Gerasopoulos, A.~Priller, C.~Schnabel, A.~Stohl, C.~Zerefos,
  H.~G\"{a}ggeler, L.~Tobler, P.~Kubik, H.~Kanter, H.~Scheel, J.~Luterbacher,
  M.~Berger, An estimate of the impact of stratosphere-to-troposphere transport
  ({STT}) on the lower free tropospheric ozone over the {A}lps using
  $\mathrm{^{10}Be}$ and $\mathrm{^7Be}$ measurements, J. Geophys. Res.
  108~(D12) (2003), 8520.
\newblock \href {http://dx.doi.org/10.1029/2002JD002604}
  {\path{doi:10.1029/2002JD002604}}.

\bibitem{Todea99}
D.~Todorovic, D.~Popovic, G.~Djuric, Concentration measurements of
  $\mathrm{^7Be}$ and $\mathrm{^{137}Cs}$ in ground level air in the {B}elgrade
  {C}ity area, Environ. Internat. 25~(1) (1999) 59--66.

\bibitem{Daiea05}
S.~Daish, A.~Dale, C.~Dale, R.~May, J.~Rowe, The temporal variations of
  $\mathrm{^7Be}$, $\mathrm{^{210}Pb}$ and $\mathrm{^{210}Po}$ in air in
  {E}ngland, J. Environ. Radioact. 84~(3) (2005) 457--467.

\bibitem{Yoshimori05}
M.~Yoshimori, Production and behavior of beryllium 7 radionuclide in the upper
  atmosphere, Adv. Space Res. 36~(5) (2005) 922--926.

\bibitem{Papas09}
C.~Papastefanou, Beryllium-7 aerosols in ambient air, Aerosol Air Qual. Res.
  9~(2) (2009) 187--197.

\bibitem{Gerea01}
E.~Gerasopoulos, P.~Zanis, A.~Stohl, C.~Zerefos, C.~Papastefanou, W.~Ringer,
  L.~Tobler, S.~H\"{u}bener, H.~G\"{a}ggeler, H.~Kanter, L.~Tositti,
  S.~Sandrini, A climatology of $\mathrm{^7Be}$ at four high-altitude stations
  at the {A}lps and the {N}orthern {A}pennines, Atmos. Environ. 35~(36) (2001)
  6347--6360.

\bibitem{Ioaea14}
A.~Ioannidou, A.~Vasileiadis, D.~Melas, Time lag between the tropopause height
  and $\mathrm{^7Be}$ activity concentrations on surface air, J. Environ.
  Radioact. 129 (2014) 80--85.

\bibitem{Kulea06}
A.~Kulan, A.~Aldahan, G.~Possnert, I.~Vintersved, Distribution of
  $\mathrm{^7Be}$ in surface air of {E}urope, Atmos. Environ. 40~(21) (2006)
  3855--3868.

\bibitem{PerHol14}
B.~R. Persson, E.~Holm, $\mathrm{^7Be}$, $\mathrm{^{210}Pb}$, and
  $\mathrm{^{210}Po}$ in the surface air from the {A}rctic to {A}ntarctica, J.
  Environ. Radioact. 138 (2014) 364--374.

\bibitem{Herea15}
M.~Hern\'{a}ndez-Ceballos, G.~Cinelli, M.~M. Ferrer, T.~Tollefsen, L.~D.
  Felice, E.~Nweke, P.~Tognoli, S.~Vanzo, M.~D. Cort, A climatology of
  $\mathrm{^7Be}$ in surface air in {E}uropean {U}nion, J. Environ. Radioact.
  141 (2015) 62--70.

\bibitem{PapIoa91}
C.~Papastefanou, A.~Ioannidou, Depositional fluxes and other physical
  characteristics of atmospheric beryllium-7 in the temperate zones
  ($\mathrm{40\,^{\circ}N}$) with a dry (precipitation-free) climate, Atmos.
  Environ. 25~(10) (1991) 2335--2343.

\bibitem{IoaPap06}
A.~Ioannidou, C.~Papastefanou, Precipitation
  scavenging of 7be and 137cs radionuclides in air, J. Environ. Radioact. 85~(1) (2006) 121--136.

\bibitem{Canea95}
F.~Cannizzaro, G.~Greco, M.~Raneli, M.~C. Spitale, E.~Tomarchio, Behaviour of
  $\mathrm{^{7}Be}$ air concentration observed during a period of 13 years and
  comparison with sun activity, Nucl. Geophys. 9 (1995) 597--607.
\newblock \href {http://dx.doi.org/10.1016/0969-8086(95)00043-7}
  {\path{doi:10.1016/0969-8086(95)00043-7}}.

\bibitem{Aliea11}
N.~Ali, E.~Khan, P.~Akhter, N.~Khattak, F.~Khan, M.~Rana, The effect of air
  mass origin on the ambient concentrations of $\mathrm{^7Be}$ and
  $\mathrm{^{210}Pb}$ in {I}slamabad, {P}akistan, J. Environ. Radioact. 102~(1)
  (2011) 35--42.

\bibitem{Chaoea12}
J.~Chao, Y.~Chiu, H.~Lee, M.~Lee, Deposition of beryllium-7 in {H}sinchu,
  {T}aiwan, Appl. Radiat. Isot. 70~(2) (2012) 415--422.

\bibitem{Garea12}
F.~{Pi\~{n}ero Garc\'{i}a}, M.~{Ferro Garc\'{i}a}, M.~Azahra, $\mathrm{^7Be}$
  behaviour in the atmosphere of the city of {G}ranada {J}anuary 2005 to
  {D}ecember 2009, Atmos. Environ. 47 (2012) 84--91.

\bibitem{Todea10}
D.~Todorovic, D.~Popovic, J.~Nikolic, J.~Ajtic, Radioactivity monitoring in
  ground level air in {B}elgrade urban area, Radiat. Prot. Dosim. 142~(2-4)
  (2010) 308--313.
\newblock \href {http://dx.doi.org/10.1093/rpd/ncq211}
  {\path{doi:10.1093/rpd/ncq211}}.

\bibitem{Ajtea08}
J.~Ajti\'{c}, D.~Todorovi\'{c}, A.~Filipovi\'{c}, J.~Nikoli\'{c}, Ground level
  air beryllium-7 and ozone in {B}elgrade, Nucl. Technol. Radiat. 23~(2) (2008)
  65--71.
\newblock \href {http://dx.doi.org/10.2298/NTRP0802065A}
  {\path{doi:10.2298/NTRP0802065A}}.

\bibitem{Ajtea13}
J.~V. Ajti\'{c}, D.~J. Todorovi\'{c}, J.~D. Nikoli\'{c}, V.~S.
  {\Dj}ur{\dj}evi\'{c}, Ground level air beryllium-7 and ozone in {B}elgrade,
  Nucl. Technol. Radiat. 28~(4) (2013) 381--388.
\newblock \href {http://dx.doi.org/10.2298/NTRP1304381A}
  {\path{doi:10.2298/NTRP1304381A}}.

\bibitem{Papea11}
S.~M. Papandreou, M.~I. Savva, K.~L. Karfopoulos, D.~J. Karangelos, M.~J.
  Anagnostakis, S.~E. Simopoulos, Monitoring of $\mathrm{^7Be}$ atmospheric
  activity concentration using short term measurements, Nucl. Technol. Radiat.
  26~(2) (2011) 101--109.
\newblock \href {http://dx.doi.org/10.2298/NTRP1102101P}
  {\path{doi:10.2298/NTRP1102101P}}.

\bibitem{Carvalea13}
A.~C. Carvalho, M.~Reis, L.~Silva, M.~J. Madruga, A decade of $\mathrm{^7Be}$
  and $\mathrm{^{210}Pb}$ activity in surface aerosols measured over the
  {W}estern {I}berian {P}eninsula, Atmos. Environ. 67~(0) (2013) 193--202.

\bibitem{Lozea13}
R.~Lozano, M.~Hern\'{a}ndez-Ceballos, J.~Rodrigo, E.~S. Miguel, M.~Casas-Ruiz,
  R.~Garc\'{i}a-Tenorio, J.~Bol\'{i}var, Mesoscale behavior of $\mathrm{^7Be}$
  and $\mathrm{^{210}Pb}$ in superficial air along the {G}ulf of {C}adiz (south
  of {I}berian {P}eninsula), Atmos. Environ. 80 (2013) 75--84.

\bibitem{Tosea14}
L.~Tositti, E.~Brattich, G.~Cinelli, D.~Baldacci, 12 years of $\mathrm{^7Be}$
  and $\mathrm{^{210}Pb}$ in {M}t. {C}imone, and their correlation with
  meteorological parameters, Atmos. Environ. 87 (2014) 108--122.

\bibitem{WalGut81}
J.~M. Wallace, D.~S. Gutzler, Teleconnections
  in the Geopotential Height Field during the Northern Hemisphere Winter,
  Mon. Wea. Rev. 109~(4) (1981) 784--812.

\bibitem{BarLiv87}
A.~G. Barnston, R.~E. Livezey, Classification, seasonality and persistence of
  low-frequency atmospheric circulation patterns, Mon. Wea. Rev. 115~(6) (1987)
  1083--1126.

\bibitem{Hurrell95}
J.~W. Hurrell, Decadal trends in the north atlantic oscillation: {R}egional
  temperatures and precipitation, Science 269~(5224) (1995) 676--679.
\newblock \href {http://dx.doi.org/10.1126/science.269.5224.676}
  {\path{doi:10.1126/science.269.5224.676}}.

\bibitem{AmbEA01}
M.~H.~P. Ambaum, B.~J. Hoskins, D.~B. Stephenson, Arctic
  Oscillation or North Atlantic Oscillation?, J. Clim. 14~(16)
  (2001) 3495--3507.

\bibitem{Wallace00}
J.~M. Wallace, North atlantic oscillatiodannular mode: Two paradigms--one phenomenon , Q. J. R. Meteorol. Soc. 126~(564) (2000) 791--805.
\newblock \href {http://dx.doi.org/10.1002/qj.49712656402}
  {\path{doi:10.1002/qj.49712656402}}.

\bibitem{LeathersEA91}
D.~J. Leathers, B.~Yarnal, M.~A. Palecki, The Pacific/North American teleconnection pattern and United States climate. Part
  I: Regional temperature and precipitation associations, J. Clim.
  4~(5) (1991) 517--528.

\bibitem{LeathersEA92}
D.~J. Leathers, M.~A. Palecki, The Pacific/North American teleconnection pattern and United States climate. Part II: Temporal characteristics and index specification, J. Clim. 5~(7) (1992) 707--716.

\bibitem{ThomWall98}
D.~W.~J. Thompson, J.~M. Wallace, The Arctic Oscillation signature
  in the wintertime geopotential height and temperature fields, Geophys. Res. Lett. 25~(9) (1998) 1297--1300.
\newblock \href {http://dx.doi.org/10.1029/98GL00950}
  {\path{doi:10.1029/98GL00950}}.

\bibitem{HigEA02}
R.~W. Higgins, A.~Leetmaa, V.~E. Kousky, Relationships
  between climate variability and winter temperature extremes in the United
  States, J. Clim. 15~(13) (2002) 1555--1572.

\bibitem{Hedea06}
J.~Hedfors, A.~Aldahan, A.~Kulan, G.~Possnert, K.-G. Karlsson, I.~Vintersved,
  Clouds and $\mathrm{^7Be}$: {P}erusing connections between cosmic rays and
  climate, J. Geophys. Res. 111~(D2), d02208.
\newblock \href {http://dx.doi.org/10.1029/2005JD005903}
  {\path{doi:10.1029/2005JD005903}}.

\bibitem{Leppea10}
A.-P. Lepp\"{a}nen, A.~Pacini, I.~Usoskin, A.~Aldahan, E.~Echer,
  H.~Evangelista, S.~Klemola, G.~Kovaltsov, K.~Mursula, G.~Possnert, Cosmogenic
  $\mathrm{^7Be}$ in air: {A} complex mixture of production and transport, J.
  Atmos. Sol-Terr. Phy. 72~(13) (2010) 1036--1043.

\bibitem{Leppea12}
A.-P. Lepp\"{a}nen, I.~Usoskin, G.~Kovaltsov, J.~Paatero, Cosmogenic
  $\mathrm{^7Be}$ and $\mathrm{^{22}Na}$ in {F}inland: {P}roduction, observed
  periodicities and the connection to climatic phenomena, J. Atmos. Sol-Terr.
  Phy. 74 (2012) 164--180.

\bibitem{LeppPaa13}
A.-P. Lepp\"{a}nen, J.~Paatero, $\mathrm{^7Be}$ in {F}inland during the
  1999-2001 {S}olar maximum and 2007-2009 {S}olar minimum, J. Atmos. Sol-Terr.
  Phy. 97 (2013) 1--10.

\bibitem{LamHes04}
J.-F. Lamarque, P.~G. Hess, Arctic oscillation modulation
  of the northern hemisphere spring tropospheric ozone, Geophys. Res. Lett. 31~(6) (2004), l06127.
\newblock \href {http://dx.doi.org/10.1029/2003GL019116}
  {\path{doi:10.1029/2003GL019116}}.

\bibitem{Herea16}
M.~Hern\'{a}ndez-Ceballos, G.~Cinelli, T.~Tollefsen, M.~Mar\'{i}n-Ferrer, Identification
  of airborne radioactive spatial patterns in Europe -- Feasibility study
  using Beryllium-7, J. Environ. Radioact. 155--156 (2016)
  55--62.

\bibitem{Hernandezetal16}
M.~Hern\'{a}ndez-Ceballos, E.~Brattich, G.~Cinelli, J.~Ajti\'{c}, V.~Djurdjevi\'{c}, Seasonality of $\mathrm{^7Be}$ concentrations in europe and influence of tropopause height, Tellus B 2016, 68, 29534.
\newblock \href {http://dx.doi.org/10.3402/tellusb.v68.29534}
  {\path{doi:10.3402/tellusb.v68.29534}}.
  
\bibitem{Klein_Tankea02}
A.~K. Tank, J.~Wijngaard, G.~K\"{o}nnen, R.~B\"{o}hm, G.~Demar\'{e}e,
  A.~Gocheva, M.~Mileta, S.~Pashiardis, L.~Hejkrlik, C.~Kern-Hansen, R.~Heino,
  P.~Bessemoulin, G.~M\"{u}ller-Westermeier, M.~Tzanakou, S.~Szalai,
  T.~P\'{a}lsd\'{o}ttir, D.~Fitzgerald, S.~Rubin, M.~Capaldo, M.~Maugeri,
  A.~Leitass, A.~Bukantis, R.~Aberfeld, A.~{van Engelen}, E.~Forland,
  M.~Mietus, F.~Coelho, C.~Mares, V.~Razuvaev, E.~Nieplova, T.~Cegnar,
  J.~{Antonio L\'{o}pez}, B.~Dahlstr\"{o}m, A.~Moberg, W.~Kirchhofer,
  A.~Ceylan, O.~Pachaliuk, L.~Alexander, P.~Petrovic, Daily dataset of
  20th-century surface air temperature and precipitation series for the
  {E}uropean {C}limate {A}ssessment, Int. J. Climatol. 22~(12) (2002)
  1441--1453.
\newblock \href {http://dx.doi.org/10.1002/joc.773}
  {\path{doi:10.1002/joc.773}}.

\bibitem{NCEP/NCAR96}
E.~Kalnay, M.~Kanamitsu, R.~Kistler, W.~Collins, D.~Deaven, L.~Gandin,
  M.~Iredell, S.~Saha, G.~White, J.~Woollen, Y.~Zhu, M.~Chelliah, W.~Ebisuzaki,
  W.~Higgins, J.~Janowiak, K.~C. Mo, C.~Ropelewski, J.~Wang, A.~Leetmaa,
  R.~Reynolds, roy Jenne, D.~Joseph, The {NCEP}/{NCAR} 40-year reanalysis
  project, Bull. Am. Meteorol. Soc. 77~(3) (1996) 437--471.

\bibitem{BraSte98}
M.~Bra\v{c}i\v{c}, A.~Stefanovska, Wavelet-based analysis of human blood-flow
  dynamics, Bull. Math. Biol. 60~(5) (1998) 919--935.
\newblock \href {http://dx.doi.org/10.1006/bulm.1998.0047}
  {\path{doi:10.1006/bulm.1998.0047}}.

\bibitem{Kikuea09}
S.~Kikuchi, H.~Sakurai, S.~Gunji, F.~Tokanai, Temporal variation of
  $\mathrm{^7Be}$ concentrations in atmosphere for 8 y from 2000 at {Y}amagata,
  {J}apan: solar influence on the $\mathrm{^7Be}$ time series, J. Environ.
  Radioact. 100 (2009) 515--521.
\newblock \href {http://dx.doi.org/10.1016/j.jenvrad.2009.03.017}
  {\path{doi:10.1016/j.jenvrad.2009.03.017}}.

\bibitem{TorCom98}
C.~Torrence, G.~P. Compo, A practical guide to wavelet analysis, Bull. Am.
  Meteorol. Soc. 79~(1) (1998) 61--78.

\bibitem{Lewea07}
J.~Lewalle, M.~Farge, K.~Schneider, Wavelet transforms, in: C.~Tropea,
  A.~Yarin, J.~Foss (Eds.), Handbook of Experimental Fluid Mechanics,
  Springer-Verlag, Berlin Heidelberg, 2007, pp. 1378--1398.

\bibitem{StratimirovicEA01}
D.~Stratimirovi\'{c}, S.~Milo\v{s}evi\'{c}, S.~Blesi\'{c}, M.~Ljubisavljevi\'{c}, Wavelet
  analysis of discharge dynamics of fusimotor neurons, Phys. A 291~(1-4) (2001) 13--23.

\bibitem{Carea04}
A.~Carbone, G.~Castelli, H.~Stanley, Time-dependent Hurst exponent in financial
  time series, Phys. A 344~(1-2) (2004) 267--271.

\bibitem{Conea13}
G.~Consolini, R.~{De Marco}, P.~{De Michelis}, Intermittency and
  multifractional {B}rownian character of geomagnetic time series, Nonlinear
  Proc. Geoph. 20~(4) (2013) 455--466.
\newblock \href {http://dx.doi.org/10.5194/npg-20-455-2013}
  {\path{doi:10.5194/npg-20-455-2013}}.

\bibitem{PengEA93}
C.-K. Peng, J.~Mietus, J.~Hausdorff, S.~Havlin, H.~Stanley, A.~Goldberger, Long-range
  anticorrelations and non-gaussian behavior of the heartbeat, Phys. Rev. Lett. 70~(9) (1993) 1343--1346.
\newblock \href {http://dx.doi.org/10.1103/PhysRevLett.70.1343}
  {\path{doi:10.1103/PhysRevLett.70.1343}}.

\bibitem{YamamotoEA06}
M.~Yamamoto, A.~Sakaguchi, K.~Sasaki, K.~Hirose, Y.~Igarashi, C.~K. Kim, Seasonal
  and spatial variation of atmospheric  $\mathrm{^{210}Pb}$ and $\mathrm{^7Be}$ deposition: features of
  the Japan Sea side of Japan, J. Environ. Radioact. 86~(1)
  (2006) 110--131.

\bibitem{AngKor64}
J.~Angell, J.~Korshover, Quasi-biennial variations in temperature, total ozone,
  and tropopause height, J. Atmos. Sci. 21~(5) (1964) 479--492.

\bibitem{NASA}
NASA (National Aeronautics and Space Administration), Marshall Space Flight Center,
http://solarscience.msfc.nasa.gov/predict.shtml visited on 28 June 2015.

\bibitem{Talposetal05}
S.~Talpos, N.~Rimbu, D.~Borsan, Solar forcing on the $\mathrm{^7Be}$-air concentration variability at ground level, J. Atmos. Sol.-Terr. Phys. 67~(16) (2005) 1626--1631.

\end{thebibliography}
\end{document}